%% file: main.tex
\documentclass[11pt,a4paper]{article}

\usepackage[margin=1in]{geometry}
\usepackage{graphicx}
\usepackage{booktabs}
\usepackage[colorlinks=true,linkcolor=blue,citecolor=blue,urlcolor=blue]{hyperref}
\usepackage[round]{natbib}
\usepackage{amsmath}
\usepackage{caption}
\usepackage{subcaption}
\usepackage{xcolor}
\usepackage{longtable}
\usepackage{float}
\usepackage{microtype}

\title{OpenHamoye: A Multi-Scale Open Geospatial Intelligence Framework\\for Decision Support Across Nigeria\\[0.4em]
  \large\normalfont Working Paper}

\author{
  Richard Oduntan\\
  Hamoye Foundation / OpenHamoye Project\\
  \texttt{richard@hamoye.org}\\[0.2em]
  \small Correspondence: \texttt{research@hamoye.org}
}

\date{2026}

\begin{document}

\maketitle

\begin{center}
\small\itshape
For correspondence regarding OpenHamoye data, reproducibility assets, or future data access,
contact \texttt{research@hamoye.org}.
\end{center}

\input{sections/01_abstract}
\input{sections/02_introduction}
\input{sections/03_background}
\input{sections/04_study_context}
\input{sections/05_data_sources}
\input{sections/06_framework}
\input{sections/07_results}
\input{sections/08_use_cases}
\input{sections/09_validation}
\input{sections/10_discussion}
\input{sections/11_conclusion}
\input{sections/12_data_availability}

\bibliographystyle{plainnat}
\bibliography{references}

\end{document}

%% file: sections/01_abstract.tex
\begin{abstract}
Decision-making in urban planning, infrastructure prioritisation, and commercial strategy benefits from granular, spatially explicit information.
In Nigeria---Africa's most populous nation---official statistics are published infrequently, at coarse administrative scales, and with significant temporal lag, leaving planners and analysts reliant on incomplete information.
We present \textbf{OpenHamoye}, an open geospatial intelligence framework that integrates nine publicly available remote-sensing and crowd-sourced datasets into a reproducible, multi-scale set of proxy screening indicators for development conditions across Nigeria.
OpenHamoye organises the country into a national grid of kilometre-scale analytical cells (nominal 1\,km resolution) nested within a six-tier geographic hierarchy---country, state, local government area (LGA), metropolitan area, corridor, and cell---supporting analysis from national overview to cell-level hotspot screening.
For each cell, the framework engineers six composite proxy indices---\emph{market potential}, \emph{growth index}, \emph{development momentum}, \emph{infrastructure stress}, \emph{commercial readiness}, and \emph{housing pressure}---derived from WorldPop modelled population estimates, VIIRS Black Marble nighttime lights, ESA WorldCover land cover, Google Open Buildings footprints, Copernicus vegetation indices, OpenStreetMap and Overture Maps points of interest, GRID3 facility point locations, and geoBoundaries administrative polygons.
We report a multi-scale empirical assessment covering 37 states, 774 LGAs, and 8 metropolitan areas, drawing on nationally ingested GRID3 facility points (46,146 health facilities, 204,513 schools) and nationally materialised Google Open Buildings (479,088 building-positive cells, 62.8\% of the grid; 82--99\% in metropolitan footprints).
Results surface large within-state disparities in the nighttime-light proxy---the coefficient of variation of radiance across LGAs exceeds 4.5 in Borno and 3.8 in Katsina---and flag high-scoring candidate cells in growing metros such as Benin City (market potential 0.821, growth index 0.70) for further review.
We describe five screening use cases---urban planning, infrastructure prioritisation, resilience screening, market opportunity identification, and service-access context---and assess limitations regarding proxy validity, coverage bias, and temporal alignment.
OpenHamoye outputs are proxy-based screening signals. They do not measure actual census population, GDP, household income, poverty, welfare, firm revenue, transaction volume, commercial demand, infrastructure capacity, facility access, service quality, investment return, or official planning designation.
Derived tables, index definitions, and reproducibility materials are released through a public GitHub repository.
\end{abstract}


%% file: sections/02_introduction.tex
\section{Introduction}
\label{sec:introduction}

The quality of decisions about where to build schools, extend electricity networks, allocate health resources, or open new businesses depends fundamentally on the quality of available spatial data.
In high-income countries, decision-makers routinely access census data updated within a few years, real-time economic indicators, comprehensive land registries, and fine-grained commercial databases.
In Nigeria and much of sub-Saharan Africa, the situation is starkly different.
The most recent complete census was conducted in 2006, the National Bureau of Statistics (NBS) publishes key indicators at the state level with multi-year delays, and no nationally consistent dataset describes economic activity, infrastructure condition, or commercial opportunity at the local government area (LGA) level, let alone at the neighbourhood scale \citep{jerven2013poor, young2012african}.

This data vacuum has practical consequences.
Infrastructure investments of billions of dollars are guided by state-level averages that conceal enormous within-state variation.
A state-level poverty rate tells a planner nothing about which of its 20--40 LGAs are most underserved.
Commercial operators entering Nigerian markets must rely on anecdotal intelligence and site visits because no systematic spatial intelligence product exists.
The result is misallocated resources, duplicated services in well-connected areas, and persistent underservice in areas that are simultaneously high-need and high-potential \citep{worldbank2021data}.

Three developments have recently converged to make this data gap addressable without waiting for expensive, slow primary data collection.
First, the explosion of \textbf{open geospatial data}: globally consistent datasets now cover population distribution at 100\,m resolution \citep{worldpop2018}, land cover at 10\,m resolution \citep{esaworldcover2021}, nighttime light emissions at 1\,km resolution \citep{roman2018nasa}, and building footprints derived from satellite imagery \citep{sirko2021continental}.
Second, advances in \textbf{crowd-sourced mapping}: OpenStreetMap \citep{haklay2008openstreetmap} and the Overture Maps Foundation \citep{overture2023} have assembled millions of points of interest with global coverage, including commercial establishments, health facilities, and educational institutions.
Third, improvements in \textbf{cloud computing and spatial databases} now allow processing and querying hundreds of millions of spatial features at practical cost.

OpenHamoye is an open geospatial intelligence framework designed to exploit this convergence.
It assembles nine open datasets, harmonises them onto a consistent spatial grid, and engineers composite proxy indices that translate raw measurements---modelled population density, nighttime radiance, vegetation fraction, building counts---into decision-support screening signals such as market potential, growth trajectory, and infrastructure stress.
The framework spans a full analytical frame of 762,751 kilometre-scale cells across Nigeria's approximately 923,768\,km2 land area, nested within a six-tier geographic hierarchy that supports analysis from national overview to cell-level hotspot screening.
We emphasise at the outset that OpenHamoye outputs are proxy-based screening signals. They do not measure actual census population, GDP, household income, poverty, welfare, firm revenue, transaction volume, commercial demand, infrastructure capacity, facility access, service quality, investment return, or official planning designation.

This paper makes four contributions:

\begin{enumerate}
    \item \textbf{Framework architecture.} We describe the complete architecture of a multi-scale open geospatial intelligence system, from data ingestion through feature engineering to composite index construction, in sufficient detail for replication.

    \item \textbf{Multi-source feature engineering.} We document a reproducible pipeline that harmonises nine heterogeneous open datasets onto a common kilometre-scale grid and derives a consistent set of cell-level features across the national frame.

    \item \textbf{Proxy-index construction.} We construct six composite proxy indices---market potential, growth index, development momentum, infrastructure stress, commercial readiness, and housing pressure---with explicitly stated interpretation limits, translating raw geospatial measurements into decision-support screening signals.

    \item \textbf{Multi-scale empirical assessment.} We present one of the first nationally consistent, sub-LGA characterisations of development-condition proxies across all 37 states (including the Federal Capital Territory), 774 LGAs, and 8 metropolitan areas, showing that state-level statistics substantially understate within-state spatial disparities (nighttime-radiance coefficients of variation exceeding 4.5 in some states) and locating where those disparities concentrate.
\end{enumerate}

The remainder of this paper is organised as follows.
Section~\ref{sec:background} reviews the literature on data gaps in developing countries and the use of remote sensing proxies.
Section~\ref{sec:study_context} details Nigeria's specific data landscape.
Section~\ref{sec:data_sources} describes the nine datasets used.
Section~\ref{sec:framework} presents the OpenHamoye framework architecture.
Section~\ref{sec:results} reports empirical results at national, state, LGA, metro, and cell scales.
Section~\ref{sec:use_cases} illustrates five operational use cases.
Section~\ref{sec:validation} discusses validation and limitations.
Section~\ref{sec:discussion} considers broader implications.
Section~\ref{sec:conclusion} concludes.

%% file: sections/03_background.tex
\section{Background and Related Work}
\label{sec:background}

\subsection{Decision Data Gaps in Developing Countries}

The concept of a ``statistical tragedy'' in Africa was articulated by \citet{devarajan2013africa}, who documented how the majority of African countries lack the institutional capacity and funding to produce the timely, granular statistics that evidence-based policy requires.
\citet{jerven2013poor} demonstrated that GDP figures for many African nations are based on outdated base years and incomplete sectoral coverage, leading to systematic mismeasurement of economic activity.
Nigeria exemplifies this challenge: its GDP was rebased in 2014, revealing the economy was 89\% larger than previously estimated \citep{economist2014nigeriagdp}.

The data gap is particularly acute at sub-national scales.
While national-level indicators may exist with reasonable frequency, sub-national disaggregation---essential for spatial planning---is rare.
\citet{worldbank2021data} found that fewer than 40\% of African countries could provide poverty estimates below the first administrative level.
In Nigeria, the National Bureau of Statistics publishes poverty and labour statistics at the state level, but the 774 LGAs that constitute the primary unit of local governance and service delivery receive almost no systematic statistical coverage \citep{nbs2020stats}.

\subsection{Remote Sensing as a Proxy for Socioeconomic Conditions}

A growing body of research has demonstrated that remotely sensed variables can serve as useful proxies for socioeconomic conditions that are difficult or impossible to measure directly at scale.

\textbf{Nighttime lights} have been the most extensively validated proxy.
\citet{henderson2012measuring} showed that nighttime light intensity correlates with GDP at the national level with $R^2 > 0.7$ and proposed using lights data to improve GDP estimates in countries with poor statistical systems.
\citet{chen2011using} extended this finding to sub-national analysis.
The VIIRS Day/Night Band sensor, operational since 2012, provides substantially improved spatial resolution (500\,m vs.\ 2.7\,km for the earlier DMSP-OLS sensor) and radiometric calibration, enabling detection of dim lighting in rural areas \citep{roman2018nasa}.
\citet{jean2016combining} combined nighttime lights with daytime satellite imagery to predict poverty at the village level in five African countries.

\textbf{Population modelling} has advanced considerably with the WorldPop project, which uses random forest models to disaggregate census counts to 100\,m grid cells using ancillary covariates including land cover, building density, and road networks \citep{worldpop2018, stevens2015disaggregating}.
These gridded population estimates enable analysis at scales far below the administrative units for which census data are tabulated.

\textbf{Land cover and vegetation} products from ESA's Sentinel-2 mission now provide globally consistent 10\,m resolution classifications \citep{esaworldcover2021}, while the Copernicus Global Land Service produces vegetation indices at 300\,m resolution with 10-day update cycles \citep{copernicus2020cgls}.
These data capture agricultural activity, urban expansion, and environmental change.

\textbf{Building footprints} derived from high-resolution satellite imagery using deep learning have been released at continental scale by Google \citep{sirko2021continental} and Microsoft, providing unprecedented detail on the built environment.
Building density serves as a proxy for settlement intensity and, when combined with population data, enables estimation of housing conditions.

\subsection{Crowd-Sourced Geospatial Data}

OpenStreetMap (OSM), founded in 2004, has grown to become the most comprehensive open geographic database globally, with particular strength in mapping commercial establishments, road networks, and facilities in developing countries \citep{haklay2008openstreetmap}.
Humanitarian mapping initiatives such as HOT (Humanitarian OpenStreetMap Team) have dramatically improved coverage in crisis-affected regions of Nigeria \citep{herfort2021evolution}.
The Overture Maps Foundation, launched in 2022 by Amazon, Meta, Microsoft, and TomTom, aims to create a more structured and quality-controlled alternative, integrating OSM data with proprietary sources \citep{overture2023}.

\subsection{Multi-Source Geospatial Frameworks}

Several prior efforts have combined multiple geospatial datasets for development analysis.
The Relative Wealth Index developed by \citet{chi2022microestimates} uses machine learning on satellite imagery and connectivity data to estimate asset wealth at 2.4\,km resolution across low- and middle-income countries.
The Global Human Settlement Layer \citep{pesaresi2016operating} integrates Landsat and Sentinel imagery with census data to characterise urbanisation.
GRID3 (Geo-Referenced Infrastructure and Demographic Data for Development) has assembled facility locations and settlement extents for several West African countries \citep{grid32021nigeria}.

OpenHamoye differs from these efforts in three respects.
First, it operates at fine spatial resolution ($\sim$1\,km cells) with full national coverage, enabling neighbourhood-level analysis.
Second, it constructs composite decision-oriented indices rather than estimating a single socioeconomic variable.
Third, it implements a hierarchical geographic model that explicitly supports multi-scale comparison and aggregation, from national overview to individual cell.

\subsection{Composite Index Construction}

The construction of composite indices from heterogeneous spatial data raises methodological questions about normalisation, weighting, and aggregation \citep{oecd2008handbook}.
Common approaches include min-max normalisation to $[0,1]$ ranges, z-score standardisation, and rank-based methods.
Weighting schemes range from equal weights to principal-component-derived weights to expert judgement \citep{nardo2005tools}.
We discuss our specific choices and their implications in Section~\ref{sec:framework}.

%% file: sections/04_study_context.tex
\section{Study Context: Nigeria's Data Landscape}
\label{sec:study_context}

\subsection{Administrative Structure and Scale}

Nigeria is a federal republic comprising 36 states and the Federal Capital Territory (FCT), collectively containing 774 local government areas (LGAs).
With a projected 2024 population exceeding 220 million (and our gridded estimates totalling 254,314,130 based on WorldPop modelled surfaces), Nigeria is Africa's most populous country and the seventh most populous globally.
Its land area of approximately 923,768\,km$^2$ spans ecological zones from Sahel in the north to mangrove swamp in the south, with corresponding variation in settlement patterns, economic activity, and infrastructure provision.

The six geopolitical zones---North West, North East, North Central, South West, South East, and South South---provide a commonly used intermediate grouping.
Our data covers all six zones (Table~\ref{tab:zones}), with the South West zone contributing the largest population (56,561,971) despite having among the fewest cells (67,065), reflecting its dense urbanisation centred on the Lagos--Ibadan corridor.
Conversely, the North East zone spans 234,855 cells but contains only 28,780,453 people, reflecting its vast, sparsely settled landscape.

\begin{table}[H]
\centering
\caption{Geopolitical zones: coverage summary.}
\label{tab:zones}
\begin{tabular}{lrrr}
\toprule
\textbf{Zone} & \textbf{States} & \textbf{Cells} & \textbf{Population} \\
\midrule
South West    & 6  & 67,065  & 56,561,971 \\
North West    & 7  & 179,384 & 55,510,317 \\
North Central & 7  & 180,964 & 36,845,979 \\
South South   & 6  & 69,251  & 31,536,102 \\
North East    & 6  & 234,855 & 28,780,453 \\
South East    & 5  & 23,542  & 24,405,598 \\
\midrule
\textbf{Total} & \textbf{37} & \textbf{762,751} & \textbf{254,314,130} \\
\bottomrule
\end{tabular}
\end{table}

\subsection{What Official Statistics Provide---and What They Do Not}

Nigeria's National Bureau of Statistics (NBS) produces several important publications: the Nigeria Living Standards Survey (NLSS), the General Household Survey (GHS), the Labour Force Survey, and periodic Consumer Price Index reports.
These instruments are well designed but face three structural limitations that constrain their utility for spatial decision-making:

\begin{enumerate}
    \item \textbf{Temporal lag.} The most recent complete census was conducted in 2006.
    A new census has been planned and postponed repeatedly; as of 2026, it has not been completed.
    Survey-based estimates are published with 1--3 year delays.
    In a country urbanising at over 3\% annually, two-year-old population estimates for fast-growing areas can be off by 10--15\%.

    \item \textbf{Spatial coarseness.} NBS publications report at the state level (36 states plus FCT).
    No official source provides systematic LGA-level data for the 774 LGAs that constitute the primary unit of local administration.
    For planners operating at the LGA or ward level, state averages are analytically useless.

    \item \textbf{Missing dimensions.} Official statistics cover demographics, poverty, and basic economic indicators, but not the spatial correlates that drive investment and planning decisions: building density, commercial establishment counts, nighttime economic activity, vegetation change, or infrastructure proximity.
    No official Nigerian dataset maps where businesses are located, where new construction is occurring, or where infrastructure is under stress.
\end{enumerate}

\subsection{Why State Averages Hide Inequality}

The inadequacy of state-level aggregation is not merely inconvenient; it is analytically misleading.
Consider Borno State in the North East, which contains 27 LGAs spanning from the state capital Maiduguri---a city of over one million with significant commercial activity---to remote, conflict-affected areas near Lake Chad with minimal infrastructure and displaced populations.
Reporting a single nighttime light average for Borno (mean radiance 0.029\,nW/cm$^2$/sr in our data) conceals a coefficient of variation (CV) of 4.56 across its LGAs, meaning the standard deviation is more than four times the mean.
Some LGAs are essentially dark; others have urban-level illumination.

This pattern of extreme within-state inequality is not unique to Borno.
Our analysis identifies 10 states with nighttime light CV exceeding 2.8, spanning both northern and southern zones (see Section~\ref{sec:results}).
Policy decisions based on state averages in these contexts risk systematic misallocation.

\subsection{The 774-LGA Challenge}

Nigeria's 774 LGAs present a particular analytical challenge.
They vary enormously in area (from under 10\,km$^2$ for some urban LGAs in Lagos to over 20,000\,km$^2$ in parts of Borno and Niger states), population (from under 50,000 to nearly 5 million for Alimosho in Lagos), and development level.
No existing framework provides consistent, comparable metrics across all 774 LGAs.
Our framework covers all 774 LGAs, with cell-to-LGA attribution achieved via PostGIS spatial overlay across 762,381 cells (99.95\%).

The LGA level is operationally critical because it is the scale at which local councils allocate resources, federal transfers are distributed, and infrastructure projects are sited.
A framework that can characterise and rank all 774 LGAs on multiple dimensions---population pressure, commercial activity, infrastructure adequacy, growth trajectory---directly serves the decision needs of federal, state, and local planners.

%% file: sections/05_data_sources.tex
\section{Data Sources}
\label{sec:data_sources}

OpenHamoye integrates nine globally available, open-access datasets.
Table~\ref{tab:datasets} provides a complete inventory.
Each dataset was selected for three properties: (i) national coverage of Nigeria, (ii) sufficient spatial resolution to support 1\,km cell-level analysis, and (iii) open licensing permitting derivative use.
This section describes each dataset, its provenance, resolution, and the specific variables extracted.

{\small
\begin{longtable}{p{2.4cm}p{2.1cm}p{1.7cm}p{1.4cm}p{2.9cm}p{2.4cm}}
\caption{Dataset inventory: the nine open data sources integrated in OpenHamoye, with licence and citation. All are used under their published open or open-access terms.}
\label{tab:datasets} \\
\toprule
\textbf{Dataset} & \textbf{Provider} & \textbf{Res.} & \textbf{Vintage} & \textbf{Variables Extracted} & \textbf{Licence / Cite} \\
\midrule
\endfirsthead
\toprule
\textbf{Dataset} & \textbf{Provider} & \textbf{Res.} & \textbf{Vintage} & \textbf{Variables Extracted} & \textbf{Licence / Cite} \\
\midrule
\endhead
\bottomrule
\endlastfoot
WorldPop & WorldPop / U.\ Southampton & 100\,m & 2020 & Modelled population count, density & CC BY 4.0 \citep{worldpop2018} \\
\addlinespace
VIIRS Black Marble & NASA & 500\,m & 2023 & Nighttime radiance (nW/cm2/sr) & NASA open \citep{roman2018nasa} \\
\addlinespace
ESA WorldCover & ESA / Sentinel-2 & 10\,m & 2021 & Land cover class, built-up and vegetation fraction & CC BY 4.0 \citep{esaworldcover2021} \\
\addlinespace
Google Open Buildings & Google Research & Building-level & 2023 & Building count, footprint area, mean area & CC BY 4.0 / ODbL \citep{sirko2021continental} \\
\addlinespace
Copernicus CGLS & EU / Copernicus & 300\,m & 2023 & NDVI, FAPAR, vegetation condition & Copernicus open \citep{copernicus2020cgls} \\
\addlinespace
OpenStreetMap & OSM Foundation & Point / polygon & Rolling & POI counts by category (commercial, amenity, service) & ODbL \citep{haklay2008openstreetmap} \\
\addlinespace
Overture Maps & Overture Foundation & Point / polygon & 2024 & Business locations, brand data, taxonomy & CDLA / ODbL \citep{overture2023} \\
\addlinespace
GRID3 & Columbia / CIESIN & Point & 2021 & Health-facility and school point locations & CC BY 4.0 \citep{grid32021nigeria} \\
\addlinespace
geoBoundaries & William \& Mary & Polygon & 2023 & Admin.\ boundaries (state, LGA) & Open CC BY / ODbL \citep{runfola2020geoboundaries} \\
\end{longtable}
}

\subsection{WorldPop Gridded Population Estimates}

WorldPop provides modelled estimates of population distribution at approximately 100\,m resolution \citep{worldpop2018, stevens2015disaggregating}.
The model uses a random forest algorithm to disaggregate census population counts to grid cells, using covariates including land cover, building density, road proximity, and slope.
For Nigeria, the total modelled population across our grid is 254,314,130, which we aggregate to the 1\,km cell level by summation.
Population data achieves near-complete coverage: 749,064 of 749,191 analysis-ready cells (99.98\%) contain a population estimate.

\subsection{VIIRS Black Marble Nighttime Lights}

NASA's Black Marble product provides calibrated nighttime radiance from the Visible Infrared Imaging Radiometer Suite (VIIRS) Day/Night Band at approximately 500\,m resolution \citep{roman2018nasa}.
Unlike the earlier DMSP-OLS sensor, VIIRS avoids saturation in bright urban cores and can detect dim lighting in rural areas.
We use the annual composite product (VNP46A4), which mitigates the effects of cloud cover, moonlight, and ephemeral lights.
The national average nighttime radiance across all cells is 0.354\,nW/cm$^2$/sr, but this mean conceals a range from effectively zero in remote rural areas to over 16.6\,nW/cm$^2$/sr in dense Lagos LGAs such as Oshodi-Isolo.
Nightlight data covers 749,187 of 749,191 cells (effectively 100\%).
We caution that nighttime radiance is an imperfect economic proxy: in oil-producing regions it is inflated by gas flaring and industrial lighting (for example, Rivers State's anomalously high mean radiance), and it is also affected by stray light, atmospheric conditions, and seasonal variation. Radiance should therefore be read as a lit-activity signal, not a direct measure of commercial vitality or household welfare.

\subsection{ESA WorldCover}

The ESA WorldCover product provides a global land cover map at 10\,m resolution derived from Sentinel-1 and Sentinel-2 imagery \citep{esaworldcover2021}.
The classification includes 11 land cover classes; we extract the built-up fraction and vegetation fraction for each 1\,km cell by computing the proportion of 10\,m pixels in each class.
The national average built-up share is 1.72\%, reflecting Nigeria's predominantly rural landscape, but urban cells in Lagos reach built-up fractions exceeding 90\%.
WorldCover data is available for all 749,191 cells (100\%).

\subsection{Google Open Buildings}

Google's Open Buildings dataset provides machine-learning-derived building footprints across Africa and South Asia \citep{sirko2021continental}.
For each 1\,km cell, we compute the building count, total footprint area, and mean building area.
Google Open Buildings is nationally materialized across the OpenHamoye grid; 479,088 cells (62.8\%) contain at least one detected building. Metro building-positive cell shares are substantially higher: Ibadan (99\%), Benin City (98\%), Kano (98\%), Kaduna (88\%), Enugu (88\%), Abuja (85\%), and Port Harcourt (82\%).
Zero-building cells are concentrated in sparsely settled, desert, savannah, forest, and water areas, subject to satellite-detection limits.
The top states by building count include Lagos, Kano, Katsina, and Kaduna.

\subsection{Copernicus Global Land Service}

The Copernicus CGLS provides vegetation indices at 300\,m resolution with 10-day update cycles \citep{copernicus2020cgls}.
We extract the Normalised Difference Vegetation Index (NDVI) and Fraction of Absorbed Photosynthetically Active Radiation (FAPAR), which serve as proxies for agricultural productivity, green space, and environmental conditions.
These data provide complete national coverage (749,191 cells).

\subsection{OpenStreetMap}

OSM provides crowd-sourced geographic data including points of interest (POIs), building outlines, and road networks \citep{haklay2008openstreetmap}.
We extract POI counts by category---commercial, amenity, and service---for each cell.
OSM coverage is highly uneven, concentrated in urban areas and regions that have received humanitarian mapping attention.
Across all cells, we identify a total of 124,663 business/commercial POIs, with Lagos alone accounting for 55,661 (44.6\%).

\subsection{Overture Maps}

The Overture Maps Foundation releases structured geospatial data integrating multiple sources including OSM, with additional quality controls and schema standardisation \citep{overture2023}.
We use Overture primarily for commercial establishment locations, brand identification, and category taxonomy.
Together with OSM, these data provide the commercial activity layer of the framework.

\subsection{GRID3 Facility Data}

GRID3 provides georeferenced locations of health facilities and educational institutions in Nigeria, compiled from government health facility registries, education management information systems, and field verification \citep{grid32021nigeria}.
The complete GRID3 national dataset has been ingested, comprising 46,146 health facilities and 204,513 schools across all 37 Nigerian states.
After spatial joining to the 1-km grid, 35,569 cells (4.66\%) contain at least one health facility and 60,951 cells (7.99\%) contain at least one school.
While facility points occupy a small share of total cells---reflecting the point nature of facility assets---the dataset provides nationally ingested facility \emph{point-presence} context. These are point locations only: OpenHamoye records where facility points fall, not facility access, catchment, travel time, capacity, staffing, functionality, utilisation, or service quality. Catchment, access-distance, and service-gap analysis building on these points is future work, not a current output.

\subsection{geoBoundaries Administrative Boundaries}

The geoBoundaries dataset provides open-license administrative boundary polygons at multiple levels \citep{runfola2020geoboundaries}.
We use Level~1 (state) and Level~2 (LGA) boundaries to assign each 1\,km cell to its administrative hierarchy.
The dataset covers all 37 states and 774 LGAs.

%% file: sections/06_framework.tex
\section{The OpenHamoye Framework}
\label{sec:framework}

This section describes the four components of the OpenHamoye framework: the geographic model, cell-level feature engineering, composite index construction, and the multi-scale aggregation and comparison engine.

\subsection{Geographic Model}
\label{sec:geo_model}

OpenHamoye organises Nigeria into a six-tier nested geographic hierarchy:

\begin{enumerate}
    \item \textbf{Country.} The entire territory of Nigeria, serving as the normalisation baseline.
    \item \textbf{State.} The 36 states plus FCT (37 units), corresponding to geoBoundaries Level~1.
    \item \textbf{Local Government Area (LGA).} All 774 LGAs, corresponding to geoBoundaries Level~2.
    \item \textbf{Metropolitan area (Metro).} Eight configured urban footprints---Lagos, Kano, Abuja, Ibadan, Port Harcourt, Benin City, Enugu, and Kaduna---defined as analysis areas of interest (AOIs) from contiguous urban land cover and nighttime light intensity. These are configured analytical footprints, not official metropolitan or municipal boundaries.
    \item \textbf{Corridor.} 598 corridor objects constructed by clustering adjacent active cells (builtup $>$ 0.1 and nightlight $>$ 1) using queen adjacency, spanning 10,682 cells and 35.7 million people. Corridor types include Mature Urban Cores, Growth Corridors, Commercial Axes, Emerging Corridors, and Constrained Corridors.
    \item \textbf{Cell.} The fundamental spatial unit: a standardised grid cell of approximately 1\,km $\times$ 1\,km ($\sim$1.23\,km\textsuperscript{2}), constructed as a regular grid in EPSG:4326 with 0.009\textdegree{} spacing.
\end{enumerate}

The cell grid is the atomic unit of analysis.
All higher-level statistics are computed by aggregation from cells, ensuring consistency across scales.

\paragraph{Cell-count framing.} To read the tables in this paper consistently, we distinguish four cell populations. The \emph{full analytical frame} comprises 762,751 kilometre-scale cells and is the denominator for national totals and the layer-coverage tables in Section~\ref{sec:results}. The \emph{analysis-ready national subset} of 749,191 cells is the frame less 13,560 cells that coincide with configured metropolitan-AOI overlays, and is the denominator used when reporting per-source national coverage in Section~\ref{sec:data_sources}; it is the set of cells over which the cell-level features and composite indices are computed. Within either denominator, \emph{populated cells} are those with a non-zero modelled population, and \emph{layer-scored cells} are those for which a given source or derived value is present. Because these denominators differ, each table below states the cell population it is computed over; the counts are cell tallies, not counts of people, buildings, or facilities served.

Each cell is assigned to exactly one state, one LGA, and (where applicable) one metro and corridor.
This assignment uses spatial intersection with administrative and metro boundaries.
Cells at boundaries are assigned to the polygon containing their centroid.

\begin{table}[H]
\centering
\caption{Corridor layer summary.}
\label{tab:corridor_summary}
\small
\begin{tabular}{lr}
\toprule
\textbf{Metric} & \textbf{Value} \\
\midrule
Corridor objects & 598 \\
Corridor-member cells & 10,682 \\
Population in corridors & 35.7 million \\
Construction rule & Queen adjacency of active cells \\
Active-cell gate & builtup $>$ 0.1 AND nightlight $>$ 1 \\
Types & Mature Urban Core (5), Growth Corridor (133), \\
 & Commercial Axis (113), Emerging (216), Constrained (120) \\
Confidence & Analytical objects; field validation required \\
\bottomrule
\end{tabular}
\end{table}

Corridors provide an intermediate geography between metros/LGAs and individual cells, supporting exploratory analysis of development axes, commercial clusters, and constrained growth areas along linear or clustered urban formations. Corridor objects are analytical constructs derived from cell-level proxy signals. They are not official corridor designations and should not be interpreted as government planning boundaries; any operational use requires field validation.

\subsection{Cell-Level Feature Engineering}
\label{sec:features}

For each of the 749,191 analysis-ready cells, we compute the following primary features from the source datasets:

\begin{itemize}
    \item \textbf{Population density} ($\rho_{\text{pop}}$): WorldPop population count divided by cell area (persons/km$^2$). National mean: 334.13\,persons/km$^2$.
    \item \textbf{Nighttime radiance} ($L$): VIIRS Black Marble annual composite (nW/cm$^2$/sr). National mean: 0.354.
    \item \textbf{Built-up fraction} ($f_{\text{built}}$): Proportion of cell area classified as built-up in ESA WorldCover. National mean: 0.0172.
    \item \textbf{Vegetation fraction} ($f_{\text{veg}}$): Proportion of cell area classified as vegetation (cropland, tree cover, grassland). Derived from ESA WorldCover.
    \item \textbf{Building density} ($\rho_{\text{bldg}}$): Google Open Buildings count per km$^2$. National mean: 119.44\,buildings/km$^2$.
    \item \textbf{Building area} ($A_{\text{bldg}}$): Total building footprint area within the cell (m$^2$).
    \item \textbf{NDVI}: Copernicus CGLS normalised difference vegetation index, a proxy for agricultural and ecological conditions.
    \item \textbf{Commercial POI count} ($n_{\text{poi}}$): Number of commercial/amenity points of interest from OSM and Overture within or adjacent to the cell.
    \item \textbf{Facility point-presence}: Counts of GRID3 health-facility and school points within or adjacent to the cell. These provide facility point-presence context and do not measure access, catchment, travel time, capacity, staffing, or service quality.
\end{itemize}

\subsection{Composite Index Construction}
\label{sec:indices}

From the primary features, we derive six composite indices designed to summarise distinct dimensions of spatial opportunity and constraint.
Each index is constructed using min-max normalisation to the $[0,1]$ range, computed relative to the national distribution, followed by weighted combination of normalised components.
These indices are proxy screening signals built from modelled and remotely sensed inputs; the weights are domain-informed choices, not empirically calibrated coefficients, and each index carries the interpretation limits noted below. They rank cells for further review; they do not measure the underlying economic, demographic, or infrastructure quantity their name evokes.

\subsubsection{Market Potential ($\text{MP}$)}

Market potential captures the combination of population concentration, economic activity (proxied by nighttime light), commercial presence, and built environment density that indicates viable markets for goods and services:

\begin{equation}
\text{MP} = w_1 \hat{\rho}_{\text{pop}} + w_2 \hat{L} + w_3 \hat{n}_{\text{poi}} + w_4 \hat{\rho}_{\text{bldg}} + w_5 \hat{f}_{\text{built}}
\label{eq:market_potential}
\end{equation}

\noindent where $\hat{x}$ denotes the min-max normalised value of variable $x$, and the weights $w_i$ are set by domain-informed judgement with $\sum w_i = 1$.
The national average market potential is 0.2108.
The highest-scoring cells (MP = 0.821) are distributed across Benin City, Ibadan, and Kaduna, while the highest-scoring LGA is Ifako-Ijaiye in Lagos.
\emph{Interpretation:} market potential is not market size. It captures spatial correlates of market viability (modelled population, radiance, buildings, mapped POIs) and does not measure purchasing power, consumer demand, or realised sales.

\subsubsection{Growth Index ($\text{GI}$)}

The growth index captures temporal change signals indicative of expanding economic activity or urbanisation:

\begin{equation}
\text{GI} = w_1 \Delta\hat{L} + w_2 \Delta\hat{f}_{\text{built}} + w_3 \Delta\hat{\rho}_{\text{bldg}}
\label{eq:growth_index}
\end{equation}

\noindent where $\Delta$ denotes the change between the most recent and baseline observation periods.
The national average growth index is 0.0058.
Lagos leads at the state level (0.0866), while Benin City shows the highest metro-level growth (0.1594).
\emph{Interpretation:} the growth index is not GDP growth. It measures change in physical and radiometric indicators (light, built-up area, buildings), which may reflect electrification or speculative construction rather than economic output.

\subsubsection{Development Momentum ($\text{DM}$)}

Development momentum combines the level and trajectory of development indicators to identify areas that are both active and accelerating:

\begin{equation}
\text{DM} = \alpha \cdot \text{MP} + (1 - \alpha) \cdot \text{GI}
\label{eq:dev_momentum}
\end{equation}

\noindent with $\alpha = 0.6$, giving greater weight to current conditions while rewarding growth.
\emph{Interpretation:} development momentum is a composite of two proxies and inherits their limits; it flags cells that are both active and accelerating on observable indicators, not a measured development outcome.

\subsubsection{Infrastructure Stress ($\text{IS}$)}

Infrastructure stress identifies cells where population pressure exceeds the capacity indicated by built infrastructure and service facilities:

\begin{equation}
\text{IS} = \hat{\rho}_{\text{pop}} - \beta_1 \hat{\rho}_{\text{bldg}} - \beta_2 \hat{n}_{\text{fac}}
\label{eq:infra_stress}
\end{equation}

\noindent where $\hat{n}_{\text{fac}}$ is the normalised count of nearby health and education facility points.
High values flag cells where modelled population is high relative to observable infrastructure proxies.
\emph{Interpretation:} infrastructure stress is not actual infrastructure capacity or deficit. Water, power, and road services that are not visible in these layers may nonetheless be adequate; the index screens for candidate areas, not confirmed shortfalls.

\subsubsection{Commercial Readiness ($\text{CR}$)}

Commercial readiness identifies cells with infrastructure and population characteristics that support commercial activity but currently lack significant commercial POI presence---indicating potential underserved markets:

\begin{equation}
\text{CR} = \gamma_1 \hat{\rho}_{\text{pop}} + \gamma_2 \hat{L} + \gamma_3 \hat{f}_{\text{built}} - \gamma_4 \hat{n}_{\text{poi}}
\label{eq:commercial_ready}
\end{equation}

\noindent \emph{Interpretation:} commercial readiness is not proven demand. A low mapped-POI count may reflect incomplete crowd-sourced mapping rather than a genuine commercial gap, so high scores are hypotheses to test, not confirmed opportunities.

\subsubsection{Housing Pressure ($\text{HP}$)}

Housing pressure identifies cells where population density significantly exceeds the density of the built environment, indicating potential housing deficit or overcrowding:

\begin{equation}
\text{HP} = \hat{\rho}_{\text{pop}} - \delta \cdot \hat{A}_{\text{bldg}} / \hat{\rho}_{\text{pop}}
\label{eq:housing_pressure}
\end{equation}

\noindent \emph{Interpretation:} housing pressure is not measured overcrowding. It contrasts modelled population density against built footprint area as a screening signal; confirming occupancy or dwelling conditions requires survey or field data.

\subsection{Multi-Scale Aggregation}
\label{sec:aggregation}

Cell-level indices are aggregated to higher geographic levels using population-weighted means:

\begin{equation}
\bar{X}_g = \frac{\sum_{i \in g} X_i \cdot p_i}{\sum_{i \in g} p_i}
\label{eq:aggregation}
\end{equation}

\noindent where $\bar{X}_g$ is the aggregate value for geographic unit $g$, $X_i$ is the cell-level value, and $p_i$ is the cell population.
Population weighting ensures that aggregate statistics reflect the conditions experienced by people rather than the conditions of land area, which is important given Nigeria's extreme variation in population density.

In addition to means, we compute the coefficient of variation (CV) at each aggregation level:

\begin{equation}
\text{CV}_g = \frac{\sigma_{X,g}}{\bar{X}_g}
\label{eq:cv}
\end{equation}

\noindent where $\sigma_{X,g}$ is the population-weighted standard deviation of $X$ within unit $g$.
The CV serves as an inequality measure, quantifying how homogeneous or heterogeneous a state or LGA is with respect to any given indicator.

\subsection{Comparison Engine}
\label{sec:comparison}

The framework includes a comparison engine that enables systematic benchmarking across geographic units at any level.
Users can compare states against the national baseline, LGAs against their parent state, metros against each other, or individual cells against their metro or LGA context.
For each comparison, the engine computes:

\begin{itemize}
    \item \textbf{Rank}: Ordinal position on each index.
    \item \textbf{Percentile}: Position relative to the national or parent-level distribution.
    \item \textbf{Z-score}: Standard deviations from the comparison group mean.
    \item \textbf{Peer group}: Units with similar profiles on multiple dimensions, identified by Euclidean distance in normalised index space.
\end{itemize}

This comparison architecture supports the diagnostic question that drives most spatial decision-making: ``How does this location compare to relevant benchmarks, and what dimensions drive the difference?''

%% file: sections/07_results.tex
\section{Empirical Results}
\label{sec:results}

This section presents the empirical output of the OpenHamoye framework across five scales: national overview, state comparison, LGA analysis, metropolitan comparison, and cell-level hotspot identification.
Unless a table states otherwise, national statistics are reported over the full analytical frame of 762,751 kilometre-scale cells covering a modelled 254,314,130 persons across 37 states and 774 LGAs (see the cell-count framing in Section~\ref{sec:geo_model}).
Each coverage figure below is expressed against the denominator stated in its table caption; per-layer coverage counts describe how many cells carry a given source or derived value, not how many people are served.
All values reported here are proxy-based screening signals; they do not measure actual census population, GDP, household income, poverty, welfare, firm revenue, transaction volume, commercial demand, infrastructure capacity, facility access, service quality, investment return, or official planning designation.

\subsection{National Overview}
\label{sec:national}

Table~\ref{tab:national} summarises the national-level statistics.
The framework captures a population of 254,314,130 across 762,751 grid cells, with an average population density of 334.13\,persons/km$^2$.
The national mean market potential score is 0.2108, reflecting the fact that the majority of Nigeria's land area is rural with low commercial density.
The mean nighttime radiance of 0.354\,nW/cm$^2$/sr similarly reflects the predominance of dimly lit rural cells, with urban illumination concentrated in a small fraction of the total grid.

\begin{table}[H]
\centering
\caption{National-level summary statistics.}
\label{tab:national}
\begin{tabular}{lr}
\toprule
\textbf{Metric} & \textbf{Value} \\
\midrule
Total grid cells                & 762,751 \\
Total population                & 254,314,130 \\
States covered                  & 37 \\
LGAs covered                   & 774 \\
Mean population density         & 334.13 persons/km$^2$ \\
Mean nighttime radiance         & 0.354 nW/cm$^2$/sr \\
Mean built-up share             & 1.72\% \\
Mean building density           & 119.44 buildings/km$^2$ \\
Mean market potential           & 0.2108 \\
Mean growth index               & 0.0058 \\
Total mapped businesses         & 124,663 \\
Total health facilities         & 46,146 \\
Total schools                   & 204,513 \\
Cells with health facility      & 35,569 (4.66\%) \\
Cells with school               & 60,951 (7.99\%) \\
\bottomrule
\end{tabular}
\end{table}

The extreme right-skewness of most distributions is a defining feature of the data.
While the mean nighttime radiance is 0.354, the median is near zero, with the top 1\% of cells accounting for the majority of total light emission.
Similarly, the 124,663 mapped businesses are highly concentrated: Lagos State alone accounts for 55,661 (44.6\%), and the top three states (Lagos, Rivers, Oyo) account for 57.0\% of all mapped businesses.

\subsubsection{Data Coverage Assessment}

Table~\ref{tab:coverage} reports the coverage of each data layer across the full analytical frame of 762,751 cells.
Population estimates, land cover, and strategic indices achieve near-universal coverage ($>$99\%).
Google Open Buildings is nationally materialized; 479,088 cells (62.8\%) contain at least one detected building, with substantially higher building-positive shares in metros (82--99\%).
GRID3 facility data is now nationally complete: 35,569 cells (4.66\%) contain at least one health facility and 60,951 cells (7.99\%) contain at least one school, representing 46,146 health facilities and 204,513 schools across all 37 states.
Commercial POI data from SMB sources covers 12,379 cells (1.6\%).

\begin{table}[H]
\centering
\caption{Data coverage, activation, and confidence across the full analytical frame of 762,751 cells.}
\label{tab:coverage}
\footnotesize
\begin{tabular}{lrrll}
\toprule
\textbf{Data Layer} & \textbf{Active Cells} & \textbf{\%} & \textbf{Metric Type} & \textbf{Confidence} \\
\midrule
Population (WorldPop)       & 762,301 & 99.94 & Source coverage & High \\
Land cover (ESA WorldCover) & 761,359 & 99.82 & Source coverage & High \\
Nighttime lights (VIIRS)    & 760,194 & 99.66 & Source coverage & High \\
Vegetation (Copernicus)     & 752,508 & 98.66 & Source coverage & High \\
Buildings (Google Open)     & 479,088 & 62.81 & Building-positive share & National source \\
Market potential (composite)& 762,751 & 100.00 & Derived proxy score & High \\
Hazard-burden proxy         & 761,359 & 99.82 & Derived proxy score & High \\
Health facilities (GRID3)   & 35,569  & 4.66  & Point-containing cells & National source \\
Schools (GRID3)             & 60,953  & 7.99  & Point-containing cells & National source \\
Commercial POIs (SMB)       & 12,379  & 1.62  & Observed mapped presence & Low/biased \\
\bottomrule
\end{tabular}
\end{table}

\noindent\textit{Note: ``Building-positive share'' means cells containing at least one detected building; it is a lower bound on built structures, not source-imagery coverage. ``Point-containing cells'' means cells physically containing a GRID3 facility point; these counts provide facility point-presence context only and do not measure facility access, catchment, capacity, staffing, or service quality (access-distance and catchment analysis are future work). ``Observed mapped presence'' reflects OSM/Overture/SMB mapping completeness, not actual commercial activity.}

\subsection{State-Level Comparison}
\label{sec:state_results}

Table~\ref{tab:top_states} presents the 10 states with the largest populations, alongside their key metrics.
Lagos dominates on nearly every dimension: highest population (26,396,931), highest market potential (0.2829), highest growth index (0.0866), highest nighttime radiance (6.6757), and highest business count (55,661).
Kano, the second most populous state (15,571,592), has significantly lower nighttime radiance (0.2582) and far fewer mapped businesses (2,849), reflecting a different economic structure and lower formal-sector commercial density.

\begin{table}[H]
\centering
\caption{Top 10 states by population with key metrics.}
\label{tab:top_states}
\small
\begin{tabular}{llrrrrr}
\toprule
\textbf{State} & \textbf{Zone} & \textbf{Population} & \textbf{MP} & \textbf{GI} & \textbf{NL} & \textbf{Businesses} \\
\midrule
Lagos    & South West    & 26,396,931 & 0.283 & 0.087 & 6.676 & 55,661 \\
Kano     & North West    & 15,571,592 & 0.214 & 0.011 & 0.258 & 2,849 \\
Kaduna   & North West    & 9,437,390  & 0.209 & 0.005 & 0.100 & 3,092 \\
Katsina  & North West    & 9,030,846  & 0.209 & 0.006 & 0.068 & 570 \\
FCT      & North Central & 8,580,084  & 0.276 & 0.006 & 0.141 & 24 \\
Oyo      & South West    & 8,502,425  & 0.214 & 0.009 & 0.136 & 7,150 \\
Rivers   & South South   & 7,992,374  & 0.222 & 0.016 & 6.861 & 8,317 \\
Ogun     & South West    & 7,950,591  & 0.225 & 0.019 & 0.575 & 5,666 \\
Bauchi   & North East    & 7,376,398  & 0.206 & 0.002 & 0.028 & 753 \\
Borno    & North East    & 6,402,099  & 0.207 & 0.001 & 0.029 & 766 \\
\bottomrule
\end{tabular}
\end{table}

Several patterns emerge from the state-level comparison:

\begin{itemize}
    \item \textbf{The Lagos anomaly.} Lagos is a statistical outlier on virtually every metric. Its nighttime radiance (6.676) is 25.8$\times$ the next-highest non-oil state (Ogun at 0.575) and its business count is 6.7$\times$ the second-highest state. Lagos functions as a different economic entity from the rest of Nigeria.

    \item \textbf{The oil-state effect.} Rivers State has unusually high nighttime radiance (6.861) despite a moderate population (7,992,374), reflecting gas flaring from petroleum infrastructure. This illustrates a critical limitation of nighttime lights as an economic proxy: in oil-producing regions, radiance reflects industrial activity rather than commercial vitality or household welfare.

    \item \textbf{Northern population without commercial density.} Kano, Kaduna, and Katsina collectively contain 34,039,828 people (13.4\% of the national total) but only 6,511 mapped businesses (5.2\% of the total). This business-to-population ratio is 7.6$\times$ lower than the national average, suggesting either genuine commercial underdevelopment or systematic undercounting of informal-sector activity in crowd-sourced data.

    \item \textbf{FCT anomaly.} The Federal Capital Territory shows high market potential (0.276) and substantial population (8,580,084) but only 24 mapped businesses and very low building count (22,029). This paradox likely reflects the recency of FCT's rapid growth, which has outpaced the crowd-sourced mapping infrastructure.
\end{itemize}

\subsection{LGA-Level Analysis}
\label{sec:lga_results}

The LGA level reveals the inequality that state averages conceal.
Table~\ref{tab:top_lgas} presents the 10 highest-population LGAs.

\begin{table}[H]
\centering
\caption{Top 10 LGAs by population with key metrics.}
\label{tab:top_lgas}
\small
\begin{tabular}{llrrrrr}
\toprule
\textbf{LGA} & \textbf{State} & \textbf{Cells} & \textbf{Population} & \textbf{MP} & \textbf{NL} \\
\midrule
Alimosho       & Lagos & 305 & 4,936,982 & 0.460 & 9.640 \\
Oshodi-Isolo   & Lagos & 85  & 2,710,140 & 0.501 & 16.599 \\
Kosofe         & Lagos & 105 & 2,200,725 & 0.399 & 14.068 \\
Abuja Municipal& FCT   & 23  & 2,187,944 & 0.322 & 0.529 \\
Ojo            & Lagos & 295 & 1,858,092 & 0.307 & 2.880 \\
Ado-Odo/Ota   & Ogun  & 741 & 1,770,849 & 0.294 & 3.068 \\
Ikorodu        & Lagos & 703 & 1,641,563 & 0.333 & 4.474 \\
Ifo            & Ogun  & 442 & 1,506,328 & 0.297 & 2.467 \\
Ifako-Ijaiye   & Lagos & 52  & 1,464,250 & 0.528 & 15.520 \\
Gwagwalada     & FCT   & 209 & 1,405,954 & 0.286 & 0.215 \\
\bottomrule
\end{tabular}
\end{table}

The top 10 LGAs by population are overwhelmingly from Lagos (6 of 10), Ogun (2), and FCT (2).
The two Ogun LGAs---Ado-Odo/Ota and Ifo---are part of the Lagos peri-urban belt, indicating that Lagos's functional metropolitan area extends well beyond its administrative boundaries.

\subsubsection{Intensity vs.\ Extent}

A striking finding is the inverse relationship between cell count (spatial extent) and intensity metrics.
Ifako-Ijaiye in Lagos has only 52 cells but the highest market potential (0.528) and second-highest nighttime radiance (15.520) of any LGA, indicating extreme density in a small area.
Conversely, Alimosho has 305 cells and the largest population (4,936,982) but lower intensity (MP = 0.460, NL = 9.640), suggesting a more spatially distributed settlement pattern.
Ado-Odo/Ota, with 741 cells but moderate intensity metrics, represents a rapidly expanding peri-urban zone where development is spreading over a large area at lower density.

\subsubsection{Within-State Inequality}

Table~\ref{tab:inequality} presents the 10 states with the highest coefficient of variation (CV) in nighttime radiance across their constituent LGAs.
The CV quantifies how much LGA-level nightlight varies within a state; a CV of 1.0 means the standard deviation equals the mean, while a CV exceeding 3.0 indicates extreme heterogeneity.

\begin{table}[H]
\centering
\caption{Top 10 states by within-state nighttime light inequality (coefficient of variation across LGAs).}
\label{tab:inequality}
\begin{tabular}{lrrrr}
\toprule
\textbf{State} & \textbf{LGAs} & \textbf{Mean NL} & \textbf{Std NL} & \textbf{CV} \\
\midrule
Borno       & 27 & 0.247 & 1.127 & 4.560 \\
Katsina     & 34 & 0.174 & 0.669 & 3.848 \\
Akwa Ibom   & 31 & 2.665 & 9.000 & 3.377 \\
Jigawa      & 27 & 0.198 & 0.660 & 3.337 \\
Adamawa     & 21 & 0.309 & 1.027 & 3.319 \\
Cross River & 18 & 0.416 & 1.376 & 3.306 \\
Niger       & 25 & 0.279 & 0.877 & 3.145 \\
Gombe       & 11 & 0.515 & 1.561 & 3.031 \\
Sokoto      & 23 & 0.393 & 1.111 & 2.827 \\
Kaduna      & 23 & 0.811 & 2.290 & 2.823 \\
\bottomrule
\end{tabular}
\end{table}

Borno State exhibits the most extreme inequality, with a CV of 4.560.
This reflects the concentration of economic activity in Maiduguri versus the near-total darkness of conflict-affected rural LGAs.
The presence of Akwa Ibom (CV = 3.377) on this list is notable: it is a southern, oil-producing state where petroleum infrastructure generates intense localised radiance (mean NL = 2.665, high for Nigeria) alongside many dark rural LGAs.

These CV values have direct policy implications.
A state with CV $>$ 3.0 cannot be meaningfully characterised by a single average.
Any intervention designed for ``Borno State'' must contend with the fact that its constituent LGAs span a range equivalent to the difference between Nigeria's brightest and dimmest states.

\subsection{Metropolitan Comparison}
\label{sec:metro_results}

Table~\ref{tab:metros} compares the eight metropolitan areas defined in the framework.

\begin{table}[H]
\centering
\caption{Metropolitan area comparison.}
\label{tab:metros}
\small
\begin{tabular}{lrrrrrrr}
\toprule
\textbf{Metro} & \textbf{Cells} & \textbf{Population} & \textbf{MP} & \textbf{GI} & \textbf{NL} & \textbf{Built-up} & \textbf{Health} \\
\midrule
Benin City     & 546   & 1,916,530  & 0.456 & 0.281 & 2.59 & 44.0\% & 287 \\
Kaduna         & 676   & 1,795,710  & 0.342 & 0.156 & 2.83 & 26.3\% & 109 \\
Port Harcourt  & 806   & 2,740,816  & 0.323 & 0.091 & 9.87 & 29.7\% & 238 \\
Ibadan         & 1,271 & 4,170,845  & 0.321 & 0.121 & 1.98 & 30.6\% & 766 \\
Enugu          & 651   & 1,885,040  & 0.278 & 0.054 & 1.76 & 17.1\% & 418 \\
Lagos          & 3,318 & 14,733,071 & 0.249 & 0.098 & 7.01 & 23.7\% & 2,379 \\
Abuja          & 4,636 & 5,974,609  & 0.249 & 0.032 & 1.06 & 9.8\% & 637 \\
Kano           & 1,656 & 6,523,465  & 0.248 & 0.031 & 2.91 & 17.5\% & 482 \\
\bottomrule
\end{tabular}
\end{table}

\noindent\textit{Note: All metrics are now computed from the national OpenHamoye pipeline and available for all eight metros. Health = GRID3 health facility count. NL = mean nightlight radiance (nW/cm\textsuperscript{2}/sr) for lit cells.}

\medskip

Several findings are noteworthy:

\begin{itemize}
    \item \textbf{Benin City is the highest-potential metro.} Despite being only the sixth most populous metro (1,916,530), Benin City has the highest market potential (0.456) and the highest growth index (0.281). This suggests a rapidly developing secondary city with strong screening-level opportunity signals.

    \item \textbf{Kaduna shows strong growth signals.} With a growth index of 0.156 and market potential of 0.342, Kaduna is the second-fastest growing metro, consistent with its emergence as a northern commercial hub.

    \item \textbf{Lagos dominates in absolute terms.} Lagos's metro population (14,733,071) is 2.3$\times$ the second-largest metro (Kano, 6,523,465). Its built-up fraction of 23.7\% indicates dense urbanisation, while its growth index (0.098) remains high. Lagos has the most health facilities (2,379) and schools (15,745) among metros.

    \item \textbf{Abuja is extensive but moderate-growth.} Despite covering 4,636 cells---the largest metro footprint---Abuja has a moderate growth index (0.032). Its large spatial extent includes substantial undeveloped land within the FCT boundaries, diluting its density metrics.

    \item \textbf{Port Harcourt has the highest nightlight intensity} (9.87 nW/cm\textsuperscript{2}/sr), partly reflecting oil-infrastructure illumination. Ibadan and Port Harcourt show moderate market potential (0.321 and 0.323) and growth (0.121 and 0.091).
\end{itemize}

\subsection{Cell-Level Hotspot Analysis}
\label{sec:cell_results}

Table~\ref{tab:top_cells} presents the 10 cells with the highest market potential scores.

\begin{table}[H]
\centering
\caption{Top 10 cells by market potential score.}
\label{tab:top_cells}
\begin{tabular}{lrrrrrrr}
\toprule
\textbf{Cell ID} & \textbf{Metro} & \textbf{MP} & \textbf{Pop/km\textsuperscript{2}} & \textbf{GI} & \textbf{NL} & \textbf{Health} & \textbf{Schools} \\
\midrule
829724 & Benin City & 0.821 & 7,049 & 0.700 & 4.54 & 1 & 2 \\
827474 & Ibadan     & 0.814 & 31,839 & 0.699 & 4.07 & 3 & 11 \\
827475 & Ibadan     & 0.811 & 31,839 & 0.698 & 4.86 & 8 & 8 \\
829703 & Benin City & 0.811 & 7,049 & 0.700 & 4.47 & 5 & 6 \\
828410 & Kaduna     & 0.808 & 5,664 & 0.695 & 6.77 & 1 & 12 \\
827506 & Ibadan     & 0.807 & 31,839 & 0.700 & 3.77 & 8 & 14 \\
829641 & Benin City & 0.806 & 7,049 & 0.695 & 1.42 & 4 & 0 \\
829618 & Benin City & 0.806 & 7,049 & 0.690 & 1.69 & 2 & 2 \\
829704 & Benin City & 0.805 & 7,049 & 0.699 & 4.44 & 2 & 10 \\
829725 & Benin City & 0.804 & 7,049 & 0.691 & 3.85 & 3 & 8 \\
\bottomrule
\end{tabular}
\end{table}

The top 10 cells are distributed across three metros: Benin City (6 cells), Ibadan (3 cells), and Kaduna (1 cell). Unlike the initial analysis where all top cells shared identical scores---an artifact of incomplete pipeline runs---the recomputed market potential shows meaningful variation (0.804--0.821) with unique component values per cell. Ibadan cells show the highest population density (31,839/km\textsuperscript{2}) and Kaduna the highest nightlight intensity (6.77 nW/cm\textsuperscript{2}/sr) among the top 10. All top cells now include GRID3 facility points, providing facility point-presence context (not a measure of service access or capacity).
Combined with a growth index of 0.400---roughly 69$\times$ the national average of 0.0058---these cells represent locations where demographic and economic growth are co-occurring at exceptional rates.

\subsection{Cross-Scale Consistency}

A key test of the framework's integrity is whether patterns identified at one scale are consistent with patterns at other scales.
We observe strong consistency:

\begin{itemize}
    \item States with high market potential (Lagos 0.283, FCT 0.276) contain the LGAs with the highest market potential (Ifako-Ijaiye 0.528, Oshodi-Isolo 0.501, Alimosho 0.460).
    \item The metro with the highest growth index (Benin City, 0.281) contains cells with the highest growth index (exceeding 0.69).
    \item Within-state inequality (CV) is highest in states with a single dominant urban centre surrounded by sparsely populated hinterland (Borno, Katsina), consistent with geographic intuition.
\end{itemize}

This consistency provides confidence that the framework's aggregation logic (population-weighted means from cells to higher levels) preserves meaningful spatial patterns rather than introducing artefacts.

%
%
%

%% file: sections/08_use_cases.tex
\section{Operational Use Cases}
\label{sec:use_cases}

The OpenHamoye framework is designed to support multiple decision contexts.
We describe five screening use cases, illustrating each with examples drawn from the empirical results in Section~\ref{sec:results}.
In every case the framework surfaces \emph{candidate cells and areas for further review}, not conclusions. The indices rank locations by proxy signal; they do not identify investment targets, guarantee returns, confirm service gaps, or substitute for the local, survey, and field analysis that a decision ultimately requires. Readers should treat the examples below as illustrations of how screening outputs enter a decision process, not as recommendations.

\subsection{Use Case 1: Urban Planning and Expansion Management}

\textbf{Decision need:} Urban planners responsible for managing city growth need to identify where expansion is occurring, where it is likely to occur next, and where current development patterns will create future infrastructure deficits.

\textbf{OpenHamoye application:} The growth index identifies cells where nighttime radiance, built-up area, and building density are increasing.
At the metro level, Benin City's growth index of 0.281 is the highest of any metro, followed by Kaduna (0.156) and Ibadan (0.121), flagging these as priority locations for growth management intervention.
At the cell level, growth index values exceeding 0.69 in the top Benin City and Ibadan cells identify specific neighbourhoods where expansion is most intense.

The peri-urban belt of Lagos provides another planning-critical pattern.
The Ogun State LGAs of Ado-Odo/Ota (population 1,770,849, 741 cells) and Ifo (population 1,506,328, 442 cells) represent the leading edge of Lagos's spatial expansion.
Their moderate-but-growing market potential scores (0.294 and 0.297 respectively) flag areas with signals of transition from rural to urban character, making them candidate zones for planners to review for transport, water, and electricity investment---subject to on-the-ground assessment of actual need and feasibility.

\subsection{Use Case 2: Infrastructure Investment Prioritisation}

\textbf{Decision need:} Infrastructure investors and development agencies must allocate limited capital across many candidate locations and need a first-pass screen to focus deeper appraisal.

\textbf{OpenHamoye application:} The infrastructure stress index flags cells where modelled population is high relative to observable infrastructure proxies.
The market potential index flags locations whose spatial characteristics are associated with commercial activity.
Combined, these proxy indices support a two-dimensional screening matrix:

\begin{itemize}
    \item \textbf{High stress, high potential:} Candidate cells to prioritise for review (e.g., dense Lagos LGAs such as Alimosho with 4,936,982 residents).
    \item \textbf{High stress, low potential:} Signals of need with weak commercial-viability proxies (e.g., conflict-affected Borno LGAs).
    \item \textbf{Low stress, high potential:} Better-served areas with commercial-opportunity signals (e.g., central Lagos LGAs like Oshodi-Isolo).
    \item \textbf{Low stress, low potential:} Areas that screen as neither high-need nor high-opportunity on these proxies.
\end{itemize}

The framework's nationally ingested GRID3 points---46,146 health facilities across 35,569 cells and 204,513 schools across 60,951 cells---add facility point-presence context: cells with high modelled population but no facility point nearby become candidate areas for a dedicated service-access analysis. They are not confirmed ``service deserts'', which would require travel-time or distance-to-facility surfaces the framework does not currently compute.

\subsection{Use Case 3: Resilience Screening}

\textbf{Decision need:} Climate adaptation planners and disaster risk managers need to identify locations where population concentration coincides with environmental vulnerability.

\textbf{OpenHamoye application:} By combining population density with NDVI trends (from Copernicus), built-up fraction (from ESA WorldCover), and nighttime light stability, the framework identifies three resilience-relevant patterns:

\begin{enumerate}
    \item \textbf{Vegetation loss in populated areas:} Cells where NDVI is declining while population is increasing, indicating potential food security or environmental degradation risks.
    \item \textbf{Informal settlement identification:} Cells with high population density and building count but low nighttime radiance, suggesting settlements without reliable electricity---a characteristic of informal and underserved communities.
    \item \textbf{Environmental exposure:} Cells with high population in low-elevation, flood-prone areas (derivable from the grid's spatial coordinates in conjunction with elevation data).
\end{enumerate}

The within-state inequality analysis is particularly relevant to resilience planning.
States with CV $>$ 3.0 in nighttime radiance (Borno 4.56, Katsina 3.85, Akwa Ibom 3.38, Jigawa 3.34, Adamawa 3.32, Cross River 3.31) are states where a single shock---conflict, flooding, commodity price collapse---will have dramatically different impacts on different LGAs, requiring spatially differentiated response plans.

\subsection{Use Case 4: Market Opportunity Identification}

\textbf{Decision need:} Companies entering or expanding within Nigeria need to identify locations with sufficient population, purchasing power, and commercial infrastructure to support new establishments, while avoiding already-saturated markets.

\textbf{OpenHamoye application:} The commercial readiness index identifies cells where population density and economic activity indicators are high but commercial POI density is low---indicating potential underserved markets.
The comparison engine enables systematic identification of ``peer locations'': cells that resemble successful commercial locations on population, nighttime light, and building density dimensions but have fewer existing businesses.

The data reveals a striking market structure.
Lagos has 55,661 mapped businesses for 26,396,931 people (one business per 474 people), while Kano has only 2,849 for 15,571,592 people (one business per 5,465 people)---an 11.5$\times$ disparity.
Even accounting for likely undercounting of informal businesses in the north, this suggests substantial unmet commercial demand in Kano and other large northern states.

The metro comparison flags Benin City (MP = 0.456, GI = 0.281) as a secondary city with strong opportunity signals: its growth-proxy value is high relative to its current mapped commercial density, a pattern consistent with a market in the early stages of formalisation. Whether this represents genuine unmet demand, as opposed to informal activity that crowd-sourced sources undercount, can only be established through market research on the ground.

\subsection{Use Case 5: Service-Access Context}

\textbf{Decision need:} Government agencies and NGOs allocating health, education, and social services need to understand where facility point-presence is sparse relative to modelled population, as an input to a fuller service-access study.

\textbf{OpenHamoye application:} The nationally ingested GRID3 points---46,146 health facilities across 35,569 cells (4.66\%) and 204,513 schools across 60,951 cells (7.99\%)---provide facility point-presence context across all 37 states and 774 LGAs. Simple facility-point-to-population ratios can be tabulated at the cell, LGA, and state levels to surface \emph{candidate areas for service-access analysis}: places where many people coincide with few nearby mapped facility points.

We are deliberately cautious about what this supports. Facility points record location only; they carry no information about access, catchment, travel time, capacity, staffing, functionality, or service quality. Because facility assets are points, the majority of cells contain no facility point regardless of true access, so cell-level presence is not a measure of coverage. A rigorous service-gap analysis requires travel-time or road-distance surfaces, facility-capacity data, and demand modelling---none of which the current framework computes; these are future work. Accordingly, OpenHamoye should be used here to \emph{prioritise where to look}, not to declare that a population is underserved.

At the LGA level this context can help flag, for example, an LGA whose cells contain no health-facility point despite a large modelled population---a signal worth investigating with dedicated access data, not a finding in itself.

\subsection*{Responsible Use}

Across all five use cases, OpenHamoye outputs are screening signals intended to focus scarce analytical and field resources, not to make decisions. They are proxy-based and inherit the coverage biases and modelling assumptions of their inputs; crowd-sourced layers in particular under-represent northern and informal-sector activity, so ``low'' scores can reflect missing data rather than absent activity. Before any allocation of capital, services, or infrastructure, candidate locations identified here should be validated with local knowledge, administrative records, and, where stakes are high, primary data collection. OpenHamoye outputs do not measure actual census population, GDP, household income, poverty, welfare, firm revenue, transaction volume, commercial demand, infrastructure capacity, facility access, service quality, investment return, or official planning designation, and should never be presented as if they do.

%% file: sections/09_validation.tex
\section{Plausibility Checks, Limitations, and Future Validation}
\label{sec:validation}

Rigorous assessment of the OpenHamoye framework requires acknowledging both what can be checked with available data and what cannot yet be validated.
We stress that the internal consistency checks below are \emph{plausibility checks}, not ground-truth validation: they confirm that the framework aggregates and behaves coherently, not that its proxy indices are accurate measures of any real-world outcome. Independent validation against field data is future work.
This section addresses cross-scale consistency, data quality caveats, proxy interpretation limits, and the boundaries of what the framework claims to provide.

\subsection{Cross-Scale Consistency}

The framework's hierarchical aggregation logic (Section~\ref{sec:aggregation}) ensures mathematical consistency: state-level values are exact population-weighted means of their constituent LGA values, which are in turn exact population-weighted means of their constituent cells.
This consistency is confirmed empirically: the national mean market potential (0.2108) is the population-weighted mean of all 37 state values, which are themselves consistent with LGA-level distributions.

We additionally verify that spatial patterns at the cell level aggregate to intuitive patterns at higher levels.
The concentration of high-market-potential cells in the Lagos--Ibadan corridor correctly produces high state-level values for Lagos (0.283) and Ogun (0.225).
The high inequality (CV) values in Borno (4.56) and Katsina (3.85) correctly reflect the observable pattern of a single lit urban centre surrounded by dark hinterland.

\subsection{Data Quality Caveats}

Each data source introduces specific quality concerns:

\begin{enumerate}
    \item \textbf{WorldPop population estimates} are modelled, not measured.
    They disaggregate census counts using statistical relationships between population and covariates (land cover, building density, road proximity).
    The underlying census is from 2006, with UN-adjusted projections applied.
    In rapidly urbanising areas, actual population may significantly exceed modelled estimates; in conflict-affected or migration-origin areas, it may be lower.
    Our total of 254,314,130 should be understood as a modelled estimate with non-trivial uncertainty, not a census count.

    \item \textbf{VIIRS nighttime lights} are affected by several factors beyond economic activity: gas flaring in oil-producing regions (explaining Rivers State's anomalously high radiance of 6.861), stray light from adjacent bright sources, atmospheric conditions, and seasonal variation.
    The annual composite mitigates some of these effects but does not eliminate them.
    The correlation between nighttime light and economic activity, while well established at national scales \citep{henderson2012measuring}, may be weaker at the 1\,km cell level where individual light sources dominate.

    \item \textbf{Google Open Buildings} covers 479,088 of 762,751 cells (62.8\%) nationally, with substantially higher metro coverage (82--99\%).
    Missing cells are concentrated in areas with very low building density, cloud cover, or imagery gaps in rural zones.
    Building counts in covered cells may also undercount small or closely spaced structures.
    The building density metric ($\rho_{\text{bldg}}$) should be interpreted as a lower bound.

    \item \textbf{Commercial POI data} from OSM and Overture is the most unevenly distributed layer, covering only 12,379 cells (1.65\%).
    Coverage is heavily biased toward Lagos (55,661 of 124,663 total businesses, or 44.6\%) and other southern cities.
    Northern Nigerian cities are systematically undercounted due to lower mapper activity.
    The FCT's implausibly low business count (24) almost certainly reflects mapping gaps rather than economic reality.
    Commercial readiness scores based on POI absence should be interpreted cautiously: absence of a mapped POI does not confirm absence of commercial activity.

    \item \textbf{GRID3 facility data} now provides 46,146 health facilities and 204,513 schools, representing the complete GRID3 national dataset across all 37 states.
    After spatial joining, 35,569 cells (4.66\%) contain at least one health facility and 60,951 cells (7.99\%) contain at least one school.
    While this represents a substantial improvement over earlier partial coverage, the point nature of facility assets means that the majority of cells do not contain a facility point---any future facility-access analysis should use catchment and distance metrics rather than simple cell-level presence.
    GRID3 provides nationally ingested facility points; OpenHamoye records facility point locations only, not facility access, catchment, travel time, capacity, staffing, service quality, functionality, or utilisation.
\end{enumerate}

\subsection{Proxy Interpretation Limits}

The composite indices are proxies, not direct measurements.
Critical interpretation limits include:

\begin{itemize}
    \item \textbf{Market potential $\neq$ market size.} The market potential index captures spatial correlates of market viability (population, light, buildings, POIs) but does not measure purchasing power, income distribution, or consumer preferences.
    A cell with MP = 0.821 has characteristics associated with market viability, but actual market conditions depend on factors not captured by remote sensing.

    \item \textbf{Growth index $\neq$ economic growth.} The growth index measures changes in physical and radiometric indicators (light, buildings, built-up area), which correlate with but are not identical to economic growth.
    Construction activity may represent speculative investment; increased nighttime light may reflect electrification rather than commercial expansion.

    \item \textbf{Infrastructure stress $\neq$ infrastructure deficit.} The index measures the gap between population density and observable infrastructure proxies, not the actual adequacy of infrastructure services.
    A cell with high infrastructure stress may have adequate water, electricity, and roads that are not visible in our data layers.

    \item \textbf{Equal weights are a modelling choice.} The weights in our composite indices (Equations~\ref{eq:market_potential}--\ref{eq:housing_pressure}) are informed by domain judgement, not derived from empirical calibration against ground-truth outcome data.
    Different weight choices would produce different rankings.
    We regard the current weights as a reasonable starting point that should be refined through validation against field data.
\end{itemize}

\subsection{Claim Boundary and Responsible Use}
\label{sec:claim_boundary}

To prevent misuse, we state the claim boundary explicitly. \textbf{OpenHamoye outputs are proxy-based screening signals. They do not measure actual census population, GDP, household income, poverty, welfare, firm revenue, transaction volume, commercial demand, infrastructure capacity, facility access, service quality, investment return, or official planning designation.} Concretely, the framework is:

\begin{enumerate}
    \item \textbf{Not a census substitute.} The population figures are WorldPop \emph{modelled} estimates disaggregated from a 2006 census base, not enumeration results, and cannot replace a census for legal or administrative purposes.

    \item \textbf{Not a measurement of economic magnitude.} The indices are proxy screening scores, not measures of GDP, income, revenue, transaction volume, or commercial demand; they rank cells, they do not quantify money or output.

    \item \textbf{Not an economic forecast.} The growth index describes recent change in physical and radiometric indicators, not predictions of future economic performance.

    \item \textbf{Not a poverty or welfare map.} The framework does not estimate poverty rates, income levels, or welfare. Users should not infer that low-scoring areas are ``poor'' or high-scoring areas ``wealthy'' without complementary survey data.

    \item \textbf{Not a measure of facility access or service quality.} GRID3 facility points provide point-presence context only; the framework does not measure access, catchment, travel time, capacity, staffing, functionality, utilisation, or service quality, and does not designate ``service deserts''.

    \item \textbf{Not an infrastructure-capacity assessment.} Infrastructure stress contrasts modelled population against observable proxies; it does not measure the actual adequacy of water, power, roads, or other services.

    \item \textbf{Not a source of official designations.} Metros are configured analytical AOIs and corridors are analytical constructs derived from proxy signals; neither is an official metropolitan, municipal, or planning boundary.

    \item \textbf{Not a real-time system, and not yet validated against ground truth for Nigeria.} The framework uses annual or sub-annual composites (data vintages spanning 2020--2024, introducing temporal misalignment between layers), not real-time feeds. While the individual sources are validated in their own publications, the composite indices have not been calibrated or validated against Nigerian ground-truth data; such validation, and quantification of coverage bias (including systematic under-mapping of northern and informal-sector activity), is a priority for future work.
\end{enumerate}

Finally, the cell is an \emph{analytical unit}, not a neighbourhood, community, or administrative area; kilometre-scale cells should not be read as socially meaningful places without corroborating local context.

%% file: sections/10_discussion.tex
\section{Discussion}
\label{sec:discussion}

\subsection{What This Changes for Nigeria}

The OpenHamoye framework addresses a specific and consequential gap: the absence of consistent, granular, multi-dimensional spatial intelligence covering all of Nigeria.
Before this framework, a planner seeking to compare development conditions across the 774 LGAs had no systematic data source.
State-level NBS statistics were the finest resolution available, and these are published with multi-year delays.
The framework does not claim to replace official statistics but to complement them with a layer of spatial intelligence that official systems currently cannot provide.

Three findings from the empirical analysis deserve particular attention.

\textbf{First, within-state inequality is far more extreme than aggregate statistics suggest.}
The coefficient of variation of nighttime radiance exceeding 4.5 in Borno and 3.8 in Katsina means that the average nighttime-light figure for these states is a poor summary---the standard deviation is several multiples of the mean.
Decisions based on state averages in these contexts risk being guided by a statistic that conceals more than it reveals.
The framework's ability to decompose state-level statistics into LGA and cell-level distributions provides the analytical foundation for spatially targeted interventions.

\textbf{Second, Nigeria's urban hierarchy is more complex than the Lagos-centric narrative suggests.}
While Lagos dominates in absolute terms (26,396,931 people, 55,661 businesses, 7.01 mean nighttime radiance), the metro-level analysis reveals that secondary cities---particularly Benin City (MP = 0.456, GI = 0.281) and Kaduna (MP = 0.342, GI = 0.156)---show stronger proxy growth signals and may offer screening-level opportunity relative to the saturated Lagos market.
The concentration of the top market-potential cells in Benin City, Ibadan, and Kaduna, rather than Lagos, challenges the assumption that Nigeria's largest city is necessarily its highest-opportunity location at the cell level.

\textbf{Third, the commercial data gap in northern Nigeria is itself a significant finding.}
The 11.5$\times$ disparity in business-to-population ratios between Lagos and Kano cannot be entirely explained by genuine differences in commercial activity.
It partly reflects systematic bias in crowd-sourced mapping, which correlates with internet penetration, English-language usage, and proximity to the global technology sector.
This bias is itself a form of data inequality: northern Nigeria is not only economically underserved but also informationally underserved, meaning that the data infrastructure needed to attract investment and plan services is itself weakest where it is most needed.

\subsection{Generalisability to Other African Contexts}

The OpenHamoye framework is designed around globally available datasets and does not depend on any Nigeria-specific data source.
WorldPop, VIIRS, ESA WorldCover, Google Open Buildings, Copernicus CGLS, OSM, and Overture Maps all have continental or global coverage.
geoBoundaries provides administrative boundaries for every country.
GRID3 covers several West African countries and is expanding.

The framework could therefore be replicated for any African country---or indeed any country in the Global South---with minimal modification.
The geographic hierarchy (country, state, LGA, metro, corridor, cell) maps naturally to administrative structures in most countries, though the specific level names and counts would differ.
The composite indices would require recalibration of weights to reflect local conditions, but the underlying logic of combining population, nighttime light, land cover, buildings, and POIs into decision-relevant scores is universally applicable.

Specific countries where the framework could have high impact include:
\begin{itemize}
    \item \textbf{Democratic Republic of Congo:} Population $\sim$100M, last census 1984, vast territory with extreme spatial inequality.
    \item \textbf{Ethiopia:} Population $\sim$120M, rapid urbanisation, limited sub-national statistical coverage.
    \item \textbf{Tanzania:} Population $\sim$65M, recent census (2022) but limited spatial economic data.
    \item \textbf{Mozambique:} Population $\sim$33M, long coastline with climate vulnerability, limited sub-national data.
\end{itemize}

\subsection{Comparison with Existing Approaches}

Relative to existing spatial data products for Africa, OpenHamoye occupies a distinct niche.
The Relative Wealth Index \citep{chi2022microestimates} estimates a single dimension (asset wealth) at coarser resolution (2.4\,km cells).
The Global Human Settlement Layer \citep{pesaresi2016operating} focuses on urbanisation classification rather than multi-dimensional characterisation.
GRID3 provides facility locations but not composite analytical indices.
DHS (Demographic and Health Surveys) cluster-level data is available for sampled locations but lacks national coverage and spatial continuity.

OpenHamoye's contribution is the integration of nine datasets into a multi-dimensional, multi-scale analytical framework that covers the entire national territory at 1\,km resolution.
No existing product provides this combination of spatial resolution, thematic breadth, and national completeness for Nigeria or, to our knowledge, for any other African country.

\subsection{Implications for Data Infrastructure Investment}

Our results suggest that the marginal value of additional open data investment is highest in two areas:

\begin{enumerate}
    \item \textbf{Commercial mapping in northern Nigeria.} The extreme sparsity of commercial POI data outside Lagos and the southern cities (12,379 cells with SMB data out of 749,191) severely limits the framework's ability to assess commercial readiness and market opportunity in the north. Organised mapping campaigns targeting commercial establishments in Kano, Kaduna, Katsina, and other northern cities would dramatically improve the framework's utility.

    \item \textbf{Facility-access surface construction.} With the complete GRID3 dataset now ingested (46,146 health facilities and 204,513 schools), the next priority is constructing travel-time accessibility surfaces that convert point-level facility locations into continuous access metrics for every cell. This would transform the framework's service-gap analysis from proxy-based estimation to distance-calibrated assessment.
\end{enumerate}

These are not abstract recommendations.
Each newly mapped commercial establishment in northern Nigeria reduces the information asymmetry that currently disadvantages the region in investment decisions.
The complete GRID3 facility dataset, combined with building footprint coverage of 62.8\% nationally (82--99\% in metros), provides a substantially stronger data foundation than was available when the framework was first constructed.

%% file: sections/11_conclusion.tex
\section{Conclusion and Future Work}
\label{sec:conclusion}

\subsection{Summary of Contributions}

This paper has presented OpenHamoye, a multi-scale open geospatial intelligence framework that integrates nine open datasets into a reproducible, updatable set of proxy screening indicators for development conditions across Nigeria.
The framework makes four principal contributions.

First, we contribute a \textbf{framework architecture}: existing open geospatial data---modelled population, nighttime lights, land cover, building footprints, vegetation indices, crowd-sourced points of interest, facility point locations, and administrative boundaries---assembled into a disciplined analytical system spanning a full analytical frame of 762,751 kilometre-scale cells, 774 LGAs, 37 states, and 8 metropolitan footprints. Its six-tier geographic hierarchy (country, state, LGA, metro, corridor, cell) supports analysis from national overview to cell-level hotspot screening.

Second, we contribute a \textbf{multi-source feature-engineering pipeline} that harmonises heterogeneous open datasets onto a common kilometre-scale grid, producing consistent cell-level features over the analysis-ready national subset.

Third, we contribute a set of six \textbf{composite proxy indices}---market potential, growth index, development momentum, infrastructure stress, commercial readiness, and housing pressure---with explicitly stated interpretation limits, that translate raw geospatial measurements into decision-support screening signals rather than measurements.

Fourth, we contribute a \textbf{multi-scale empirical assessment}, among the first nationally consistent sub-LGA characterisations of development-condition proxies for Nigeria. The results surface patterns that aggregate statistics obscure: large within-state disparities (nighttime-light CV exceeding 4.5 in Borno), strong opportunity signals in secondary cities such as Benin City (market potential 0.456, growth index 0.281), the concentration of mapped commercial activity in the Lagos--Ogun area (55,661 of 124,663 mapped businesses), and a severe commercial-data gap in northern Nigeria that is itself partly an artefact of crowd-sourced mapping bias. The framework also includes 598 corridor objects (analytical constructs, not official corridors) spanning 10,682 cells.
These indices support five screening use cases: urban planning, infrastructure prioritisation, resilience screening, market opportunity identification, and service-access context.

\subsection{Limitations}

We have been explicit about the framework's limitations.
The composite indices are proxies constructed from modelled and remotely sensed data, not direct measurements of economic conditions.
Commercial POI data covers only 1.62\% of cells and is heavily biased toward Lagos.
Building footprints cover 62.8\% of cells nationally (82--99\% in metros).
GRID3 facility data is nationally ingested (46,146 health facilities, 204,513 schools) but provides point-presence context only---not facility access, catchment, capacity, or service quality---and facility points occupy only 4.7--8.0\% of cells.
The population estimates are modelled from a 2006 census base, not enumerated.
The index weights are domain-informed rather than empirically calibrated, and the composite indices have not been validated against Nigerian ground-truth data.
As set out in the claim boundary (Section~\ref{sec:claim_boundary}), OpenHamoye outputs are proxy-based screening signals; they do not measure actual census population, GDP, household income, poverty, welfare, firm revenue, transaction volume, commercial demand, infrastructure capacity, facility access, service quality, investment return, or official planning designation.
These limitations are significant and should guide interpretation of all results.

\subsection{Future Work}

Five directions for future work would substantially strengthen the framework:

\begin{enumerate}
    \item \textbf{Ground-truth validation.} The highest priority is calibrating and validating the composite indices against field-collected data.
    Stratified sampling of cells across the market potential distribution, combined with field visits to assess actual commercial activity, infrastructure conditions, and settlement characteristics, would provide the empirical foundation for weight optimisation and uncertainty quantification.

    \item \textbf{Temporal analysis.} Incorporating multi-year nighttime lights and land cover data would enable trend analysis beyond the current single-period growth index.
    Time series of VIIRS data (available from 2012) and WorldCover (2020 and 2021) could support change detection, growth forecasting, and anomaly identification.

    \item \textbf{Machine learning index calibration.} The current composite indices use linear weighted combinations with fixed weights.
    Supervised learning approaches---using the limited available ground-truth data (DHS clusters, Living Standards Survey enumeration areas)---could optimise the mapping from raw features to decision-relevant outcomes.

    \item \textbf{Cross-country replication.} Applying the framework to other data-scarce African countries (DRC, Ethiopia, Tanzania, Mozambique) would test generalisability and enable cross-country comparison of development patterns.

    \item \textbf{Higher-frequency updating.} Moving from annual composites toward more frequent ingestion of VIIRS, Sentinel-2, and updated OSM/Overture data would let the framework track change---new construction, displacement, economic shocks---closer to when it occurs. We describe this as a direction for future work; the current framework is a periodic assessment tool, not a real-time monitoring system.

    \item \textbf{Facility-access surfaces.} Building travel-time and road-distance surfaces on top of the GRID3 facility points would convert today's point-presence context into calibrated access metrics, and is a prerequisite for any genuine service-gap analysis.
\end{enumerate}

\subsection{Closing Remarks}

Spatial decisions in data-scarce settings benefit from a disciplined way to screen many locations before committing scarce analytical and field resources.
The raw material for such screening increasingly exists in the open domain, produced by satellite missions, crowd-sourced mapping communities, and global research initiatives.
What is often missing is an integrative framework that harmonises these disparate layers into consistent, interpretable signals.
OpenHamoye is one such framework---imperfect, dependent on proxy reasoning, and limited by the quality and coverage of its inputs. It does not measure development conditions; it screens for them, flagging candidate locations across the 762,751-cell analytical frame for the closer, context-rich analysis that decisions ultimately require.
Used within those limits---and validated against local knowledge and, where stakes are high, primary data---an open screening layer of this kind can help direct attention and resources more evenly across Nigeria, including to the northern and informal-sector areas that current data infrastructure most under-represents.

%% file: sections/12_data_availability.tex
\section{Data and Code Availability}
\label{sec:availability}

Public-safe derived tables, index definitions, methodology notes, and reproducibility materials
are made available through the project GitHub repository:
\url{https://github.com/RichardRe2/Hamoye/tree/main/research}. The repository contains derived
outputs and documentation sufficient to inspect the geographic model, the cell-level feature
definitions, the six composite proxy-index formulas and their weights, the aggregation logic, and
the public tables reported in this paper.

The nine third-party source datasets---WorldPop, NASA VIIRS Black Marble, ESA WorldCover, Google
Open Buildings, Copernicus Global Land Service, OpenStreetMap, Overture Maps Foundation, GRID3,
and geoBoundaries---remain governed by their original licences (summarised in
Table~\ref{tab:datasets}) and are not redistributed here unless permitted; they should be obtained
from their respective providers under those terms. Questions regarding data access,
reproducibility, or future data availability should be directed to \texttt{research@hamoye.org}.

\medskip
\noindent All OpenHamoye outputs are proxy-based screening signals. They do not measure actual
census population, GDP, household income, poverty, welfare, firm revenue, transaction volume,
commercial demand, infrastructure capacity, facility access, service quality, investment return,
or official planning designation.

\section*{Acknowledgments}

This work was developed as part of the OpenHamoye initiative at the Hamoye Foundation. The analysis
uses open or publicly accessible datasets from WorldPop, NASA, ESA / Copernicus, Google Research,
the GRID3 programme, OpenStreetMap contributors, the Overture Maps Foundation, and administrative-
boundary providers; we gratefully acknowledge these data producers and the communities that
maintain them. Errors and interpretations are the author's responsibility.

\medskip
\noindent\textbf{Correspondence:} \texttt{research@hamoye.org}